\documentclass[aps,prl,twocolumn,showpacs,superscriptaddress]{revtex4}
\usepackage{amsmath}
\usepackage{amssymb}
\usepackage{epsfig}
\usepackage{color}
\usepackage{graphicx,amsmath}
\makeatletter
\renewcommand{\@biblabel}[1]{#1.}
\makeatother

\begin{document}
\title{Scaling behavior of the thermopower of
the archetypical heavy-fermion metal $\rm{YbRh_2Si_2}$}

\author{V. R. Shaginyan}\email{vrshag@thd.pnpi.spb.ru}
\affiliation{Petersburg Nuclear Physics Institute, NRC Kurchatov
Institute, Gatchina, 188300, Russia}\affiliation{Clark Atlanta
University, Atlanta, GA 30314, USA} \author{A. Z.
Msezane}\affiliation{Clark Atlanta University, Atlanta, GA 30314,
USA}\author{G. S. Japaridze}\affiliation{Clark Atlanta University,
Atlanta, GA 30314, USA}\author{K. G. Popov}\affiliation{Komi Science
Center, Ural Division, RAS, Syktyvkar, 167982,
Russia}\affiliation{Department of Physics, St.Petersburg State
University, Russia}
\author{J.~W.~Clark}
\affiliation{McDonnell Center for the Space Sciences \& Department
of Physics, Washington University, St.~Louis, MO 63130, USA}
\affiliation{Centro de Ci\^encias Matem\'aticas, Universidade de
Madeira, 9000-390 Funchal, Madeira, Portugal}
\author{V. A. Khodel}
\affiliation{McDonnell Center for the Space Sciences \& Department
of Physics, Washington University, St.~Louis, MO 63130,
USA}\affiliation{Russian Research Center Kurchatov Institute,
Moscow, 123182, Russia}

\begin{abstract}
We reveal and explain a scaling behavior of the thermopower
$S/T$ exhibiting by the archetypical heavy-fermion (HF) metal
$\rm{YbRh_2Si_2}$ under the application of magnetic field $B$ at
temperatures $T$. We show that the same scaling is demonstrated
by such different HF compounds as $\beta$-${\rm YbAlB_4}$ and
the strongly correlated layered cobalt oxide $\rm
[BiBa_{0.66}K_{0.36}O_{2}]CoO_{2}$. Using $\rm{YbRh_2Si_2}$ as
an example, we demonstrate that the scaling behavior of $S/T$ is
violated at the antiferromagnetic phase transition, while both
the residual resistivity $\rho_0$ and the density of states $N$
experience jumps at the phase transition, making the thermopower
experience two jumps and change its sign. Our elucidation is
based on flattening of the single-particle spectrum that
profoundly affects $\rho_0$ and $N$. To depict the main features
of the $S/T$ behavior, we construct the $T-B$ schematic phase
diagram of $\rm{YbRh_2Si_2}$. Our calculated $S/T$ for the HF
compounds are in good agreement with experimental facts and
support our observations.
\end{abstract}

\pacs{71.27.+a, 72.15.Jf, 64.70.Tg \\ Keywords: thermoelectric
and thermomagnetic effects, quantum phase transition, flat
bands, non-Fermi-liquid states, strongly correlated electron
systems, heavy fermions}

\maketitle

\section{Introduction}

It is ordinarily believed that the behavior of heavy-fermion
(HF) metals is determined by quantum critical point (QCP) that
suppresses quasiparticle excitations and generates non-Fermi
liquid (NFL) behavior, revealing vivid deviations from the
Landau Fermi liquid (LFL) behavior, see e.g. \cite{col,loh}. The
NFL behavior is commonly characterized by a set of exponents of
the temperature dependence of the physical properties, such as
specific heat, resistivity, susceptibility etc.
\cite{col,loh,shagrep,book}. The LFL and NFL behaviors, and the
crossover region cannot be captured by any single exponent as
seen, for example, from Fig. \ref{fig1} (a), that depicts the
behavior of the normalized specific heat $(C/T)_N$ extracted
from measurements of the specific heat $C/T\propto M^*$ on $\rm
YbRh_2Si_2$ under the application of magnetic fields $B$
\cite{oeschler:2008}, where $M^*$ is the effective mass and $T$
is temperature. It is seen that the curves $(C/T)_N$ obtained in
measurements at different magnetic fields $B$ merge into a
single one, exhibiting the scaling behavior
\cite{shagrep,book,shaginyan:2009}. This scaling behavior
obtains an adequate description within the framework of fermion
condensation theory (FC) that supports the extended
quasiparticle paradigm \cite{shagrep,book}.

Thermopower $S/T$ is a sensitive and helpful probe to
disentangle the electronic excitations at the Fermi surface.
Thus, we face an important problem related to the revealing of
scaling behavior of the thermopower $S/T$, that allows one to
analyze the nature of electronic excitations at the Fermi
surface. Along this line we shall clarify the role of
quasiparticles and flat bands, and the nature of electronic
excitations that form the behavior of the thermopower $S/T$ in
different HF compounds.

In this rapid communication we demonstrate that the thermopower
$S/T$ of the archetypical HF metal $\rm{YbRh_2Si_2}$ exhibits a
scaling behavior that coincides with that of other thermodynamic
functions like $(C/T)_N$. We show that $S/T$ of such different
HF compounds as $\rm{YbRh_2Si_2}$, $\beta$-${\rm YbAlB_4}$, and
$\rm [BiBa_{0.66}K_{0.36}O_{2}]CoO_{2}$ exhibits the same
scaling behavior, coinciding with that of the normalized
specific heat $(C/T)_N$ shown in Fig. \ref{fig1} (a). Using the
archetypical HF metal $\rm{YbRh_2Si_2}$ as an example, we also
demonstrate that the universal behavior of $S/T$ is violated at
the AF phase transition, while the residual resistivity $\rho_0$
and the density of states $N$ experience jumps at the phase
transition. This results in corresponding downward jumps of
$S/T$ and its change of sign. To depict the main features of the
$S/T$ behavior, we construct a schematic $T-B$ phase diagram.
Our calculated $S/T$ of ${\rm YbRh_2Si_2}$, $\beta$-${\rm
YbAlB_4}$ and $\rm [BiBa_{0.66}K_{0.36}O_{2}]CoO_{2}$ are found
to be in good agreement with experimental observations.

\section{Scaling behavior}

A study of the thermoelectric power $S/T$ may deliver new
insight into the nature of the quantum phase transition that
defines the NFL behavior of the corresponding HF compound. For
example, one may reasonably propose that the thermoelectric
power $S/T$ distinguishes between two competing scenarios for
quantum phase transitions in heavy fermions, namely the
spin-density-wave theory and the breakdown of the Kondo effect
\cite{pepin1,pepin}. Indeed, $S/T$ is sensitive to the
derivative of the density of electronic states and the change in
the relaxation time at $\mu$ \cite{abrikos,lan}. Using the
Boltzmann equation, the thermopower $S/T$ can be written as
\cite{abrikos,lan,beh,miy,zlat}
\begin{equation}
\frac{S}{T}=-\frac{\pi^2k_B^2}{3e}\left[\frac{\partial\ln\sigma(\varepsilon)}
{\partial\varepsilon}\right]_{\varepsilon=\mu},\label{SB}
\end{equation}
where $k_B$ and $e$ are, respectively, the Boltzmann constant
and the elementary charge, while $\sigma$ is the dc electric
conductivity of the system, given by
\begin{equation}
\sigma(\varepsilon)=2e^2\tau(\varepsilon)\int\delta(\mu-\varepsilon({\bf
p}))v({\bf p})v({\bf p})\frac{d{\bf p}}{(2\pi)^3},\label{SSG}
\end{equation}
${\bf p}$ is the electron wave-vector, $\tau$ is the scattering
time, and  $v$ denotes the velocity of electron. Thus, we see
from Eq. \eqref{SSG} that the thermoelectric power $S/T$ is
sensitive to the derivative of the density of electronic states
$N(\varepsilon=\mu)$ and the change in the relaxation time at
$\varepsilon=\mu$. On the basis of the Fermi liquid theory
description, the term in the brackets on the right hand side of
Eq. \eqref{SB} can be simplified, so that one has $S/T\propto
N(\varepsilon=\mu)\propto C/T\propto M^*$ at $T\to0$
\cite{lan,beh,miy,zlat}. As a result, upon taking into account
that charge and heat currents at low temperatures are
transported by quasiparticles, the ratio
\begin{equation}
(S/C)\simeq (S/S_{ent})\simeq const,\label{sent}
\end{equation}
where $S_{ent}$ the entropy density of charge carriers
\cite{lan,beh,miy,zlat}. Thus, we expect that within FC theory
one can obtain an adequate description of the thermopower and
its scaling behavior, for the theory is based on the
quasiparticle paradigm \cite{ks,noz,physrep,shagrep,book}.

\begin{figure}[!ht]
\begin{center}
\includegraphics [width=0.47\textwidth]{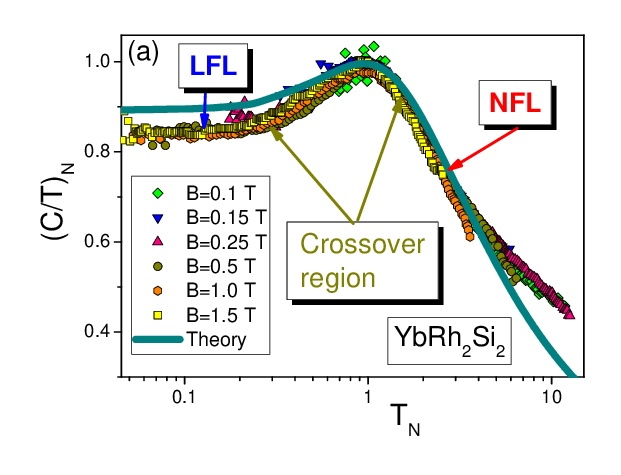}
\includegraphics [width=0.47\textwidth]{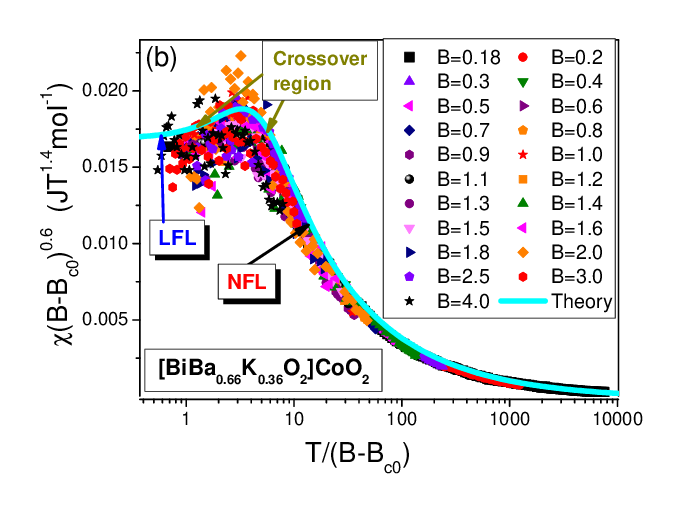}
\end{center}
\vspace*{-0.3cm} \caption{The scaling behavior of the
thermodynamic functions. (a) The normalized specific heat
$(C/T)_N$ versus normalized temperature $T_N$. $(C/T)_N$ is
extracted from the measurements of the specific heat $C/T$ on
$\mathrm{YbRh_2Si_2}$ in magnetic fields $B$\,\,
\cite{oeschler:2008} listed in the legend. The LFL region,
crossover one, and NFL one are depicted by the arrows. The solid
curve displays calculations of $(C/T)_N=M^*_N$ based on Eqs.
\eqref{HC1} and \eqref{UN2} \cite{shaginyan:2009}. (b) Scaled
susceptibility $\chi(B-B_{c0})^{0.6}$ as a function of scaled
temperature $T/(B-B_{c0})$ with $B_{c0}=0.176$ T for various $B$
values shown in the legend \cite{limel}. The LFL region,
crossover one, and NFL one are shown by the arrows. The solid
curve tracks the results of our calculations based on Eq.
\eqref{HC1} that describe the universal scaling behavior
$(C/T)_N=M^*_N\propto\chi(B-B_{c0})^{0.6}$ shown in
Fig.~\ref{fig1} (a).} \label{fig1}
\end{figure}
In the FC theory, QCP is interpreted as the fermion-condensation
quantum phase transition (FCQPT) at which the quasiparticle
effective mass $M^*$ diverges. In such an event, quasiparticles
of energy $\varepsilon$ remain well-defined excitations near the
chemical potential $\mu$, $\varepsilon\sim \mu$
\cite{shagrep,book}, while the FC state itself is protected by
topological invariants \cite{volovik,volns}. In the vicinity of
FCQPT it is helpful to use "internal" scales to measure such
quantities as e.g. $C/T$, $M^*$ and the temperature $T$ in order
to reveal the universal scaling behavior of $M^*$ observed in HF
compounds \cite{shagrep,book}. Maximum structures
$(C/T)_M\propto M^*_M$ in both $C/T$ and $M^*$, respectively, at
temperature $T_M$ appear with the application of magnetic field
$B$ and $T_M$ acquires higher values as $B$ is increased. To
obtain $(C/T)_N$, we use $(C/T)_M$ and $T_M$ as "internal"
scales: $(C/T)_M$ is used to normalize $C/T$, and $T$ is
normalized by $T_M$ \cite{shagrep,book}. In the same way we
normalize $M^*$ to obtain the normalized effective mass
$M^*_N=M^*/M^*_M$ as a function of the normalized temperature
$T_N=T/T_M$. To study the scaling behavior of $M^*(B,T)$, we use
the model of homogeneous HF liquid, that permits to avoid
complications associated with the crystalline anisotropy of
solids, while the Landau equation describing $M^*(T,B)$ of HF
liquid reads \cite{land,shagrep,book}
\begin{equation}
\frac{1}{M^*(T,B)}=\frac{1}{M}+\int\frac{{\bf p}_F{\bf
p}}{p_F^3}F({\bf p_F},{\bf p})\frac{\partial n(T,B,{\bf
p)}}{\partial{\bf p}}\frac{d{\bf p}}{(2\pi)^3}, \label{HC1}
\end{equation}
where $M$ is the corresponding bare mass, $F({\bf p_F},{\bf p})$
is the Landau interaction, which depends on Fermi momentum
$p_F$, momentum $p$, and $n$ is the distribution function. Near
FCQPT, the normalized solution of Eq. \eqref{HC1} $M^*_N(T_N)$
can be well approximated by a simple universal interpolating
function \cite{shagrep,book}. The interpolation occurs between
the LFL and NFL regimes and represents the universal scaling
behavior of $M^*_N$
\begin{equation}M^*_N(y)\approx c_0\frac{1+c_1y^2}{1+c_2y^{8/3}}.
\label{UN2}
\end{equation}
Here, $y=T_N=T/T_{M}$, $c_0=(1+c_2)/(1+c_1)$, $c_1$, $c_2$ are
fitting parameters. Magnetic field $B$ enters Eq. \eqref{HC1}
only in the combination $B/T$, making $T_{M}\propto B$. Thus, in
the presence of fixed magnetic field the variable $y$ becomes
$y=T/T_{M}\sim T/B$. Thus, Eq. \eqref{UN2} describes the
universal scaling behavior of $M^*_{N}$ as a function of $T$
versus $B$ - the curves $M^*_{N}$ at different magnetic fields
$B$ merge into a single one in terms of the normalized variable
$y=T/T_M$. In the same way, Eq. \eqref{UN2} describes the
scaling behavior of $M^*_{N}(B,T)$ as a function of $B$ versus
$T$\, \cite{land,shagrep,book,shaginyan:2009}.

Fig.~\ref{fig1} (a) shows the resulting $(C/T)_N$ as a function
of $T_N$, with different symbols for different magnetic field
strengths $B$. The solid curve represents calculations of
$(C/T)_N=M^*_N$ based on Eqs. \eqref{HC1} and \eqref{UN2}
\cite{shaginyan:2009}. It is seen that the LFL and NFL regions
are separated by a crossover region where $(C/T)_N$ reaches its
maximal value. As evident from Fig.~\ref{fig1} (a), $(C/T)_N$ is
not a constant as would be for a LFL; furthermore, it
demonstrates the asserted universal scaling behavior given by
Eq. \eqref{UN2} over a wide range of values of the normalized
temperature $T_N$. This behavior coincides with that of the
magnetic susceptibility $\chi\propto C/T\propto M^*$ revealed in
measurements on $\rm [BiBa_{0.66}K_{0.36}O_{2}]CoO_{2}$
\cite{limel} and incorporated in Fig.~\ref{fig1} (b). The solid
curve tracks the results of the same calculations  based on Eq.
\eqref{HC1} that describe the universal scaling behavior
$(C/T)_N(T/T_{M})=M^*_N(T/T_{M})\propto\chi(B-B_{c0})^{0.6}$
shown in Fig.~\ref{fig1} (a). We thus conclude that the solid
curve drawn in Figs.~\ref{fig1} (a) and (b) exhibits the
universal scaling behavior intrinsic to HF compounds
\cite{shagrep,book}.
\begin{figure}[!ht]
\begin{center}
\includegraphics [width=0.47\textwidth]{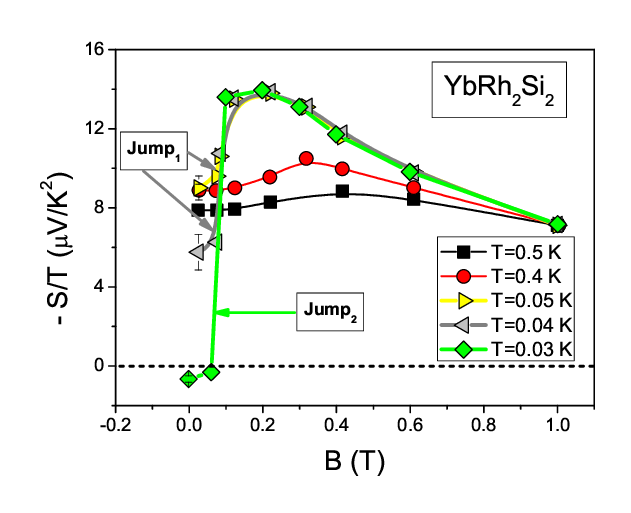}
\end{center}
\vspace*{-0.3cm} \caption{Thermopower isotherm $-S(B)/T$ for
different temperatures shown in the legend \cite{TP,TPJ}. The
labels $\rm Jump_1$ and $\rm Jump_2$ represent the first and
second downward jumps in $-S(B)/T$ shown by the arrows. The
solid lines are guides to the eye.} \label{fig2}
\end{figure}
It is seen from Fig. \ref{fig2} that in the case of
$\mathrm{YbRh_2Si_2}$ and at $T\geq T_{NL}$, the isotherms
$-S(B)/T$ behave like $C/T$: They exhibit a broad maximum that
sharpens and shifts to lower fields upon cooling \cite{TP,TPJ}.
It is seen from Fig. \ref{fig2} that the mentioned behavior is
violated as the system approaches the AF phase transition taking
place at $T_{NL}(B)$. Here $T_{NL}(B)$ is the temperature of
antiferromagnetic (AF) ordering, with $T_{NL}(B=0)=70$ mK, and
$T_{NL}(B=B_{c0})=0$ at the critical field $B_{c0}=60$ mT,
applied perpendicular to the magnetically hard c axis
\cite{shall}. Thus, we can expect that outside the AF region,
$S/T\propto C/T\propto\chi\propto M^*$ over a wide range of $T$
and $B$, since in the framework of FC theory quasiparticles are
responsible for the thermodynamic and transport properties
\cite{shagrep,book}. It is worth noting that $S/T\propto M^*$ in
a low-disorder two-dimensional electron system in silicon, and
tends to a diverge at a finite disorder-independent density
\cite{shash}.

\begin{figure}[!ht]
\begin{center}
\includegraphics [width=0.47\textwidth]{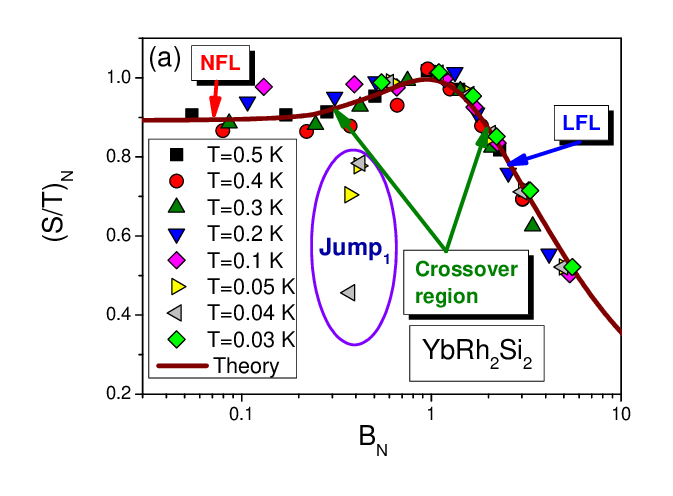}
\includegraphics [width=0.47\textwidth]{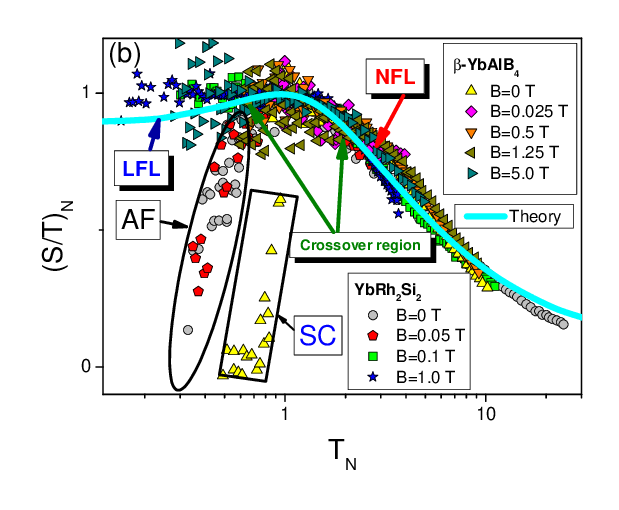}
\end{center}
\vspace*{-0.3cm} \caption{Scaling behavior of $S/T$. (a)
Normalized isotherm $(S(B)/T)_N$ versus normalized magnetic
field $B_N$ for different temperatures shown in the legend.
Outside the AF phase transition, the data follow the universal
scaling curve shown in Fig. \ref{fig1} (a). (b) Temperature
dependence of the normalized thermopower $(S/T)_N$ under several
magnetic fields shown in the legend. The experimental data are
extracted from measurements on $\rm{YbRh_2Si_2}$ \cite{TP,TPJ}
and on $\beta$-${\rm YbAlB_4}$\, \cite{machida}. As it is
explained in the text, the data, taken at the AF phase
\cite{TP,TPJ} and at the superconducting one (SC) \cite{machida}
and confined by both the ellipse and the rectangle,
respectively, signal of violation of the scaling behavior. The
solid curves in (a) and (b) represent calculated $(C/T)_N$
displayed in Fig \ref{fig1} (a) \cite{shaginyan:2009}.}
\label{fig3}
\end{figure}

To reveal the scaling behavior of the thermopower $S/T\propto
C/T\propto M^*$, we normalize $S/T$ in the same way as in the
normalization of $C/T$: the normalized function $(S/T)_N$ is
obtained by normalizing $(S/T)$ by its maximum value, occurring
at $T=T_M$, and the temperature $T$ is scaled by $T_M$. Taking
into account that $S/T\propto C/T$\, \cite{beh,miy,zlat}, we
conclude that $(S/T)_N=(C/T)_N=M^*_N$, provided that the system
in question is located away from possible phase transitions.
This function $(C/T)_N=M^*_N$ is displayed in Fig.~\ref{fig1}
(a). Figures \ref{fig3} (a) and (b) report $(S/T)_N$ as a
function of the normalized magnetic field $B_N$ and $T_N$,
respectively. In Fig. \ref{fig3} (a), the function $(S/T)_N$ is
obtained by normalizing $(S/T)$ by its maximum occurring at
$B_M$, and the field $B$ is scaled by $B_M$. As seen from Eq.
\eqref{UN2}, the LFL behavior takes place at $B_N>1$, since
$(S/T)_N=M^*_N$, and $M^*_N\propto (B-B_{c0})^{-2/3}$ are
$T$-independent, while at $B_N<1$, $M^*_M$ becomes $T$-dependent
and exhibits the NFL behavior with $M^*_N\propto T^{-2/3}_N$. It
is seen from Figs. \ref{fig3} (a) and (b) that the calculated
values of the universal function $M^*_N$ are in good agreement
with the corresponding experimental data over the wide range of
the normalized magnetic field. Thus, $(S/T)_N=(C/T)_N=M^*_N$
exhibits the universal scaling behavior over a wide range of its
scaled variable $B_N$ and $T_N$. Figure \ref{fig3} (a) also
depicts a violation of the scaling behavior for $B\leq B_{c0}$
when the system enters the AF phase. Moreover, as seen from
Figs. \ref{fig2} and \ref{fig3} (a) and (b), the scaling
behavior is violated at $T\leq T_{NL}$ by two downward jumps.
The first jump, shown in Figs.~\ref{fig2} and \ref{fig3} (a) and
labeled $\rm Jump_1$, takes place at $T_{NL}>T>0.3$ K, while the
second, occurring at $T\leq 0.03$ K, is shown in Fig.~\ref{fig2}
and labeled $\rm Jump_2$. The latter is accompanied by a change
of sign of $(S/T)_N$, which now becomes positive \cite{TP,TPJ}.
As we shall see, these two jumps reflect the presence of flat
band at $\mu$ in the single particle spectrum $\varepsilon({\bf
p})$ of heavy electrons in $\rm{YbRh_2Si_2}$\,
\cite{shagrep,book}. In the same way, as it is seen from Fig.
\ref{fig3} (b), the scaling behavior is violated by the
superconducting (SC) phase transition, taking place in
$\beta$-${\rm YbAlB_4}$ at $T_c\simeq$ 80 mK \, \cite{machida}.

\begin{figure}[!ht]
\begin{center}
\includegraphics [width=0.47\textwidth]{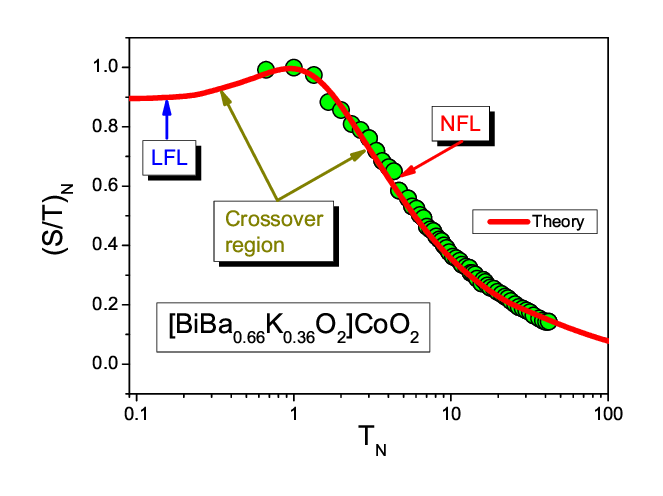}
\end{center}
\vspace*{-0.3cm} \caption{Scaling behavior of $S/T$ in the
strongly correlated layered cobalt oxide $\rm
[BiBa_{0.66}K_{0.36}O_{2}]CoO_{2}$. Temperature dependence of
$(S(T)/T)_N$ at magnetic field $B=0$, extracted from
measurements on $\rm [BiBa_{0.66}K_{0.36}O_{2}]CoO_{2}$
\cite{limel}, is displayed versus $T_N$. The solid curve
representing the theoretical calculations is the same as that
depicted in Fig. \ref{fig1} (a).} \label{fig6}
\end{figure}
We now further show that the observed scaling behavior of
$(S/T)_N$ is universal by analyzing the experimental data on the
thermopower for $\rm [BiBa_{0.66}K_{0.36}O_{2}]CoO_{2}$
\cite{limel}. The solid curve, representing our calculations in
Fig. \ref{fig6}, is the same as that depicted in Fig. \ref{fig1}
(a), and describes $(C/T)_N$ extracted from measurements on the
archetypical HF metal $\rm YbRh_2Si_2$ \cite{shaginyan:2009}. By
plotting $(S/T)_N$ as a function of $T_N$ in Fig. \ref{fig6},
the universal scaling behavior and the three regimes are seen to
be in a complete agreement with the reported overall behavior in
both Figs. \ref{fig1} (a), and Figs. \ref{fig3} (a), (b) as
well. Our preliminary result show that $\rm Ce$-based HF
compounds like $\rm CeCoIn_5$ exhibit the same scaling behavior
of $(S/T)_N$, and will be published elsewhere.

\section{Flat bands and the jumps in $S/T$ at the AF phase
transition}

\begin{figure}[!ht]
\begin{center}
\includegraphics [width=0.47\textwidth]{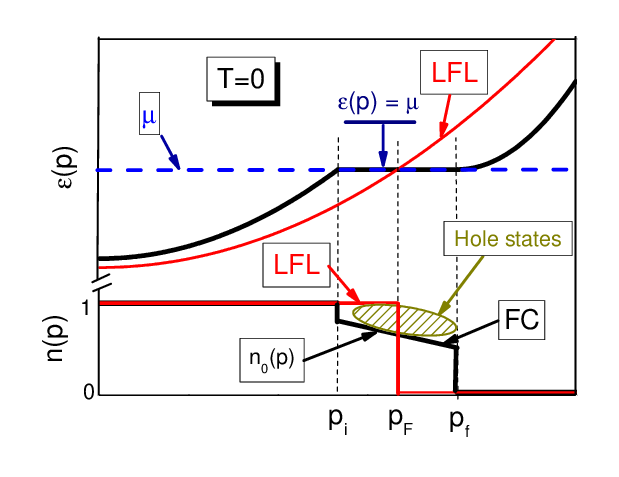}
\end{center}
\vspace*{-0.3cm} \caption{Single-particle energy
$\varepsilon({\bf p})$ and distribution function $n({\bf p})$ at
$T=0$. The arrow shows the chemical potential $\mu$. The
vertical lines show the area $p_i<p<p_f$ occupied by FC with
$0<n_0(p)<1$ and $\varepsilon({\bf p})=\mu$. The Fermi momentum
$p_F$ satisfies the condition $p_i<p_F<p_f$ and corresponds to
the LFL region, indicated by the arrows, that emerges when the
FC state is eliminated. The arrow depicts hole states induced by
the FC.} \label{fig5a}
\end{figure}
Before proceeding to the analysis of the jumps observed in
measurements of $S/T$ on $\rm{YbRh_2Si_2}$, some remarks are in
order concerning the flattening of the spectra $\varepsilon({\bf
p})$ in HF systems, a phenomenon called swelling of the Fermi
surface or FC \cite{shagrep,book,ks}. As indicated in
Fig.~\ref{fig5a}, the ground states of systems with flat bands
are degenerate, and therefore the occupation numbers $n_0({\bf
p})$ of single-particle states belonging to a flat band are
given by a continuous function on the interval $[0,1]$, in
contrast to the FL restriction to occupation numbers 0 and 1.
This property leads to an entropy excess
\begin{equation}
S_0=-\sum n_0({\bf p})\ln n_0({\bf p})+(1-n_0({\bf p}))\ln(1-
n_0({\bf p})),\label{S*}
\end{equation}
that does not contribute to the specific heat $C(T)$. The
entropy excess $S_0$ contradicts the Nernst theorem. To
circumvent violation of the Nernst theorem, FC must be
completely eliminated at $T\to 0$. This can happen by virtue of
some phase transition, e.g., the AF transition that becomes of
first order at some tricritical point occurring at $T=T_{\rm
tr}$ \cite{shagrep,book}. Such a first-order phase transition
provides for eradication of the flat portion in the spectrum
$\varepsilon({\bf p})$. As a consequence, both the density of
states $N$ and the hole states, shown by the arrow in Fig.
\ref{fig5a}, vanish discontinuously, while the occupation
numbers $n_0({\bf p})$ and the spectrum $\varepsilon({\bf p})$
revert to those LFL state, as indicated by the arrows in
Fig.~\ref{fig3} (a). Simultaneously, the Fermi sphere undergoes
an abrupt change on the interval from the Fermi momentum $p_f$
to $p_F$, so as to nullify both the swelling of the Fermi
surface and the entropy excess $S_0$. As result, the thermopower
experiences $\rm Jump_2$, as it is follow from Eq. \eqref{sent},
for the entropy abruptly diminishes. We note that the abrupt
change is observed as the change of the low-$T$ Hall coefficient
\cite{nsteg,shagrep,shall,book}. It is seen from Fig. \ref{fig2}
that at $T=0.03$ K, $S/T$ abruptly change it sign (the second
jump - $\rm Jump_2$), for the hole states vanish. The positive
sign of $S/T$ of $\rm YbRh_2Si_2$ without the hole states
\cite{miy} is in agreement with the positive thermopower of its
nonmagnetic counterpart $\rm LuRh_2Si_2$, lacking the $4f$ hole
states at $\mu$ \cite{TP,TPJ,steg8}. Contrary, at
$T_{NL}>T>T_{cr}$ the AF phase transition is of the second order
and the entropy is a continuous function at the border of the
phase transition. Therefore, at this second order phase
transition both the occupation numbers and the spectrum do not
change, and keep their FC-like shape, while the system with FC
is destroyed, converting into HF liquid. This destruction
generates the first jump $\rm Jump_1$, shown in Fig. \ref{fig2}.
Therefore, as the FC state is decayed, its contribution
$\rho_0^{FC}$ to the residual resistivity $\rho_0$ vanishes,
resulting in the change of the scattering time
$\tau(\varepsilon=\mu)$. We recall that in the presence of FC,
the residual resistivity consists of two terms
$\rho_0=\rho_0^{FC}+\rho_0^{imp}$, where the residual
resistivity $\rho_0^{FC}$ is formed by the flat band generated
by FC, while the resistivity $\rho_0^{imp}$ is formed by
impurities \cite{book,shaginyan:2012:C,shaginyan:2012:B}. As a
result, the thermopower experiences the first jump $\rm Jump_1$,
as seen from Eqs. \eqref{SB} and \eqref{SSG}. We also conclude
that the second downward jump under decreasing $B$ is deeper
than the first one, since it is caused by elimination of both
$\rho_0^{FC}$ and the hole states. This is consistent with the
experimental observations, as seen from Fig.~\ref{fig2}.

\section{Schematic $T-B$ phase diagram}

\begin{figure}[!ht]
\begin{center}
\includegraphics [width=0.47\textwidth]{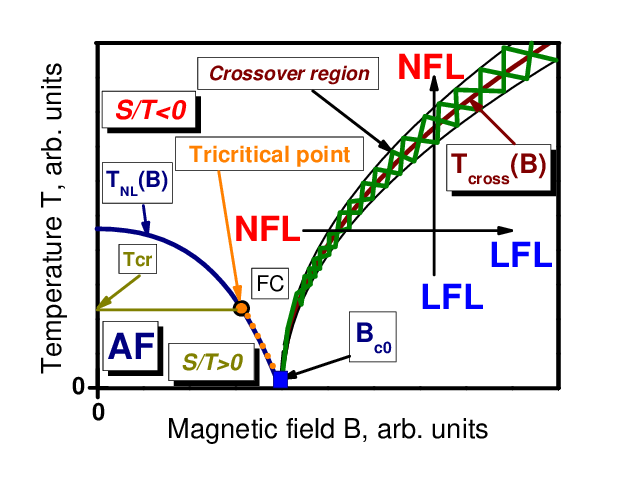}
\end{center}
\vspace*{-0.3cm} \caption{Schematic $T-B$ phase diagram of $\rm
YbRh_2Si_2$. The vertical and horizontal arrows, crossing the
transition region depicted by the thick lines, show the LFL-NFL
and NFL-LFL transitions at fixed $B$ and $T$, respectively. The
hatched area around the solid curve $T_{\rm cross}(B)$
represents the crossover between the NFL and LFL domains. The
NFL region is labeled by FC. As shown by the solid curve, at
$B<B_{c0}$ the system is in its AF state, and exhibits the LFL
behavior \cite{shall}. The line of AF phase transitions is
denoted by $T_{NL}(B)$. The tricritical point, indicated by the
arrow, is at $T=T_{\rm cr}$. At $T<T_{\rm cr}$ the AF phase
transition becomes of the first order.} \label{fig5b}
\end{figure}
Now we are in position to construct the $T-B$ phase diagram of
the archetypical HF metal $\rm{YbRh_2Si_2}$. Fig.~\ref{fig5b}
displays our constructed schematic $T-B$ phase diagram for
$\rm{YbRh_2Si_2}$. In Fig.~\ref{fig5b}, the NFL region, formed
by the FC state, is characterized by the entropy excess $S_{0}$
given by Eq.~\eqref{S*}, and labeled by FC. As shown by the
solid curve denoted by $T_{NL}(B)$, at $B<B_{c0}$ and
$T<T_{NL}(B)$ the system is in its AF state, and exhibits the
LFL behavior \cite{shall}. The tricritical point $T_{\rm cr}$ at
which the AF phase transition becomes of the first order, is
indicated by the arrow. At that transition the thermopower
experiences the jump $\rm Jump_2$ shown in Fig. \ref{fig2}, and
changes its sign, becoming $S/T>0$, for the hole states shown in
Fig. \ref{fig5a} vanish at $T<T_{\rm cr}$. At $T>T_{\rm cr}$ the
AF transition of the second order, at which the thermopower
experiences $\rm Jump_1$, see Fig. \ref{fig2}. While in the NFL
region $S/T>0$, as it is shown in the phase diagram \ref{fig5b}.
Clearly, on the basis of the phase diagram \ref{fig5b}, outside
the area of the AF phase transition, the behavior of
$S_N=M^*_N$, considered as a function of the dimensionless
variable $T_N$ or $B_N$, is almost universal. Indeed, as seen
from Figs. \ref{fig1} (a), (b), \ref{fig3} (a), (b), and
\ref{fig6}, all the data, extracted from measurements on $\rm
YbRh_2Si_2$, $\beta$-${\rm YbAlB_4}$, and $\rm
[BiBa_{0.66}K_{0.36}O_{2}]CoO_{2}$, collapse on the single
scaling curve shown in Fig. \ref{fig1} (a). As seen from
Figs.~\ref{fig1} (a) and \ref{fig3} (b), at $T_N<1$, $(S/T)_N$
tends to become constant, implying that $S/T$ exhibits LFL
behavior. However, at $T_N\simeq 1$ the system enters the narrow
crossover region, while at growing temperatures, NFL behavior
prevails.

\section{Conclusions}

In summary, we have revealed and explained the universal scaling
behavior of the thermopower $S/T$ in such different HF compounds
as $\rm{YbRh_2Si_2}$, $\beta$-${\rm YbAlB_4}$, and $\rm
[BiBa_{0.66}K_{0.36}O_{2}]CoO_{2}$. Our calculations are in good
agreement with experimental observations, and demonstrate that
the advocated universal scaling behavior of $S/T$ does take
place. This behavior does not depend on the specific properties
of the considered HF compounds, and coincides with that of the
normalized effective mass $M^*_N=(C/T)_N$, thus representing the
scaling behavior intrinsic to HF compounds. We have also shown
that destruction of the flattening of the single-particle
spectrum profoundly affects $S/T$, leading to the two jumps and
the change of sign of the thermopower occurring at the
antiferromagnetic phase transition.\\

{\bf Acknowledgements} VRS thanks the RSF Grant, \# 14-22-00281.
AZM thanks the US DOE, Division of Chemical Sciences, Office of
Energy Research, and ARO for research support. KGP is supported
by Grant \# 11.38.658.2013 and RFBR \# 14-02-00044.  VAK and JWC
acknowledge research support from the McDonnell Center for the
Space Sciences, and JWC thanks the University of Madeira
gracious hospitality during frequent visits.

\end{document}